\documentstyle{elsart}

\begin{document} 

\input{epsf.tex} 
 
\begin{frontmatter}
 
\title{PRECISE MEASUREMENT OF  {\huge $\Sigma$} BEAM ASYMMETRY
\vskip 0.0 true cm
 FOR POSITIVE PION PHOTOPRODUCTION ON THE PROTON
\vskip 0.0 true cm
 FROM 800 TO 1500 MEV }

\collab{GRAAL Collaboration}

\author[rom]{O.Bartalini}
\author[cat]{V.Bellini}
\author[isn]{J.-P.Bocquet}
\author[rom]{M.Capogni}
\author[gen]{M.Castoldi}
\author[rom]{A.D'Angelo}
\author[ipn]{J.-P.Didelez}
\author[rom]{R.Di Salvo}
\author[rom]{A.Fantini}
\author[tor]{G.Gervino}
\author[san]{F.Ghio}
\author[san]{B.Girolami}
\author[ipn]{M.Guidal}
\author[ipn]{E.Hourany}
\author[inr]{V.Kouznetsov\thanksref{slava}}
\author[ipn]{R.Kunne}
\author[inr]{A.Lapik}
\author[fra]{P.Levi Sandri}
\author[isn]{A.Lleres}
\author[rom]{D.Moricciani}
\author[inr]{V.Nedorezov}
\author[isn]{L.Nicoletti}
\author[isn]{D.Rebreyend}
\author[isn]{F.Renard}
\author[itp]{N.Roudnev}
\author[rom]{C.Schaerf}
\author[cat]{M.Sperduto}
\author[cat]{M.Sutera}
\author[kur]{A.Turinge}
\author[gen]{A.Zucchiatti}

\collab{and GW-SAID group}

\author[gwu]{R.Arndt}
\author[gwu]{W.Briscoe}
\author[gwu]{I.Strakovsky}
\author[gwu]{R.Workman\thanksref{ron}}

\address[rom]{INFN Sezione di Roma II and Universit\`a ``Tor Vergata",
00133, Italy}
\address[cat]{INFN Sezione di Catania and Universit\`a di Catania, 95100, Italy}
\address[isn]{IN2P3, Institut des Sciences Nucl\'eaires, 38026 Grenoble, France }
\address[gen]{INFN Sezione di Genova and Universit\`a di Genova, 16146, Italy }
\address[ipn]{IN2P3 Institut de Physique Nucl\'eaire, 91406 Orsay, France }
\address[tor]{INFN Sezione di Torino and Universit\`a di Torino, 10125, Italy }
\address[san]{INFN Sezione di Roma I and Instituto Superiore di Sanit\'a,
00161 Roma, Italy}
\address[inr]{Institute for Nuclear Research, 117312 Moscow, Russia }
\address[fra]{INFN Laboratori Nazionali di Frascati, 00044, Italy }
\address[itp]{Institute of Theoretical and Experimental Physics, 117259 Moscow,
Russia }
\address[kur]{RRC Kurchatov Institute, 123182 Moscow, Russia}
\address[gwu]{The George Washington University, 20052 Washington DC, USA }

\thanks[slava]{Contact person: E-mail Slava@cpc.inr.ac.ru, Slava@gravax2.in2p3.fr}
\thanks[ron]{Contact person: E-mail Ron@alpha1.phys.gwu.edu }

\date{\today}  

\begin{abstract}

The $\Sigma$ beam asymmetry for positive pion photoproduction 
on the proton has been measured over an angular range of 
$40-170^{\circ}$ at photon energies from 0.8 to 1.5~GeV.  The 
resulting data set includes 237 accurate points, 136 of these 
belonging to an almost unexplored domain above 1.05~GeV.  Data 
of such high precision provide severe constraints for partial 
wave analyses. The influence of this experiment on the GW 
multipole analysis is demonstrated.  Significant changes are
found in multipoles connected to the $S_{31}(1620)$ and 
$P_{13}(1720)$ resonances.  Comparisons using the MAID analysis are also 
presented.


\end{abstract} 



\end{frontmatter}


The spectrum of baryon resonances contains important 
information regarding the internal structure of the 
nucleon.  Precise determinations of resonance properties 
(masses, widths, and electromagnetic couplings) provide 
vital benchmarks for the development of quark models.
Much of our present knowledge has been obtained through 
pion-nucleon scattering.  The existence of many 
resonances has been established and some properties have 
been determined with reasonable precision \cite{pdg}. 
However, many features of the baryon spectrum remain a 
mystery.  Among 43 nucleon and delta resonances 
predicted by QCD-inspired models \cite{qcd}, only 24 
have been well established (``four-star" or ``three-star" 
resonances \cite{pdg}).  The remaining (existing) 
resonances are either masked by overlapping states or 
are weakly excited in reactions coupled to the 
single-pion$-$nucleon channel.

Meson photoproduction offers a complementary approach to
the baryon spectroscopy. The photoproduction of mesons other 
than pions allows access to states which could be 
suppressed in the pion-nucleon scattering.  Here, and also in 
the case of pion photoproduction, the beam asymmetry 
observable $\Sigma$ has proven to be particularly 
sensitive to resonance properties \cite{sigma}.

Properties of resonances are extracted from the 
photoproduction data by means of partial wave analysis 
and multipole decomposition in the framework of different 
approaches \cite{str,maid}.  The comparison of calculated
observables to experimental data constrains theoretical 
models \cite{LiSag,feu} and determines the role and 
properties of the included resonances.  The quality of 
this procedure is directly related to the quality of the 
data base.  The extraction of resonance parameters clearly 
requires both the unpolarized cross section and 
polarization observables \cite{tab,work}.  While the cross 
section is a source of information on dominating
components of the scattering amplitude, the polarization 
observables, in particular the polarized-photon 
beam-asymmetry $\Sigma$, are much more sensitive to the 
non-dominant contributions. 

The high yield and kinematic simplicity of single-pion 
photoproduction offers an opportunity to produce high 
precision experimental data.  Several recent attempts 
have been made to study the nucleon and Delta resonances 
using single-pion photoproduction data, from which 
parameters of the dominant resonances have been extracted 
\cite{str,maid,feu}. Uncertainties for other resonances 
remain large, due both to the quality of available data 
and model-dependence inherent in the extraction process.

The world data base is rather extensive, containing 20,000 
data points for a single-pion photoproduction \cite{str} 
below 2~GeV.  However, this count includes many old 
unpolarized cross sections measured with bremsstrahlung 
beams in the first and second resonance regions.  There is 
a clear lack of polarized data above 1~GeV.  Only a few 
tens of beam asymmetry points, of low accuracy and dated 
before 1980, are available for the $\pi^+n$ final state.
In this Letter, we report a new measurement of the beam 
asymmetry $\Sigma$ for positive-pion photoproduction on 
the proton over the energy range of 0.8$-$1.5~GeV. We 
shall see that, by adding a significant number of precise
points to the data base, an important constraint is 
placed on the existing partial wave analyses. 
 
The present data have been obtained at the GRAAL 
facility.  The polarized and tagged photon beam is 
produced by backscattering of laser light on 6.04~GeV 
electrons which circulate in the storage ring of the ESRF 
(Grenoble, France).  Through the use of green 514~nm 
laser light, the tagged spectrum covers an energy range 
of 0.55$-$1.1~GeV.  Alternately, the UV 351~nm line can be 
employed, resulting in an energy range of 0.8$-$1.5~GeV. 
The linear beam polarization varies from $\sim$0.45 at 
the lower energy limits to 0.98 at the upper limits. 
The detection system includes three main parts:\\
- At forward angles $\theta_{lab} \leq 25^{\circ} $ there 
  are two planar wire chambers, a thin time-of-flight (TOF)
 hodoscope made up of 26 horizontal and 26 vertical plastic
 scintillator strips, each 3~cm thick, and a TOF shower wall
 \cite{rw}.  The latter is an assembly of 16 modules, each
 being a sandwich of four converter-plus-scintillator layers. \\
- At central angles from 25 to 155$^{\circ} $, the target 
 is surrounded by two cylindrical wire chambers, a 5~mm thick 
 scintillator $\Delta E$ barrel, and a BGO ball made up of
 480 crystals, each of 21 radiation lengths \cite{bgo}. \\
- At backward angles $\theta_{lab} \geq 155^{\circ} $ there
 are two plastic scintillator disks separated by a 1~cm lead
 converter. \\
The apparatus provides the detection and identification 
of all types of final state particles in an almost 4$\pi$ 
solid angle and has cylindrical symmetry, making it suitable 
for measurements of the beam asymmetry $\Sigma$.  The detailed 
description of the GRAAL facility and the procedure used in 
deriving $\Sigma$ from experimental data may be found in 
\cite{rw,bgo,gra,pin}.
 
The present measurement is a high energy extension of the 
previous one, performed at GRAAL in the energy range of
0.6$-$1.1~GeV by using  the green laser line \cite{pin}. 
One important advantage of this measurement is the use of
the high resolution cylindrical chambers, in coincidence 
with the scintillator barrel and BGO ball, to detect pions 
and to reconstruct their trajectories.  As a result, the 
determination of the scattering angle $\Theta _{cm}$ is 
made possible with an  average  accuracy of $2.5^{\circ}$ 
(FWHM).  This feature allows a reduction in the widths of 
angular bins and decreases systematic uncertainties 
related to the granularity of the BGO ball.  Furthermore, 
an improved background rejection is achieved by requesting 
this triple coincidence between the chambers, the barrel, 
and the BGO ball.    

\begin{figure} 
\centerline{\epsfverbosetrue\epsfxsize=10.5 cm\epsfysize=8.5cm
\epsfbox{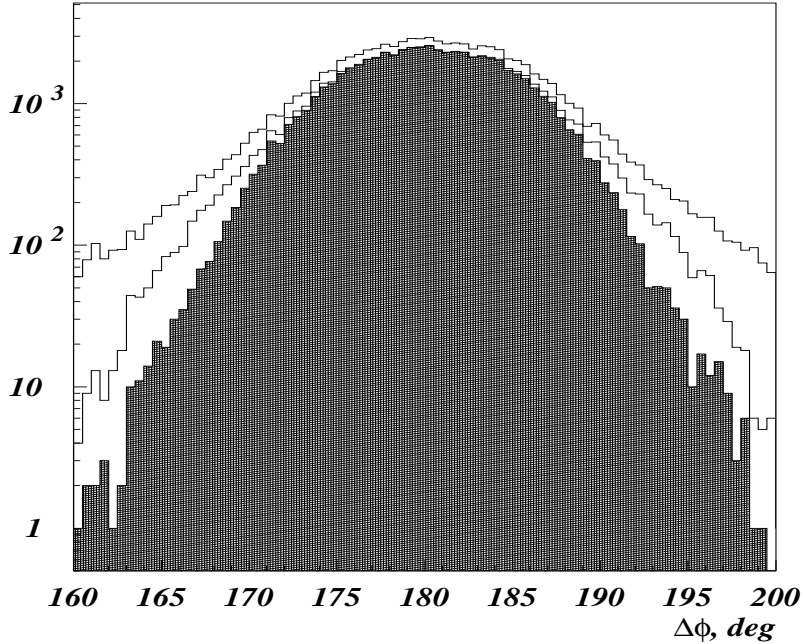}} 
\vspace*{.4cm} 
\caption{Spectrum of the coplanarity $(\phi_n-\phi_{\pi})$. Upper curve
corresponds to the selection of one neutral and one charged hit in the detector;
medium curve is after the identification of a neutron and a pion; black
area indicates finally selected events. }
\label{Figure1}
\end{figure} 

Neutrons at forward angles, corresponding to backward 
pions, were detected in the forward shower wall.  Their 
discrimination from other types of particles is made 
possible by means of the anticoincidence of the wall with 
the preceding planar chambers and scintillator hodoscope, 
and by using a $\Delta E-$TOF relation \cite{rw}.  Neutrons 
emitted at central angles above $25^{\circ}$ were detected 
in the BGO ball. The wire chambers and the scintillator 
barrel surrounding the target allow a discrimination 
between neutral and charged particles.  Further separation 
between photons and neutrons is possible, considering the 
number of BGO crystals in the cluster which correspond to 
a neutral particle hit. As found in GEANT simulations,
neutron clusters normally do not contain more than 3 
crystals.  Photons, however, produce larger clusters of a 
size depending on the deposited energy, and usually 
include 4$-$10 crystals. 
The identification of charged pions and their discrimination from 
protons is achieved by using $\Delta E-E$ relations between the energy
$\Delta E$ deposited by a charged particle in the thin scintillator 
barrel and the corresponding energy deposition $E$ in the BGO ball.

\begin{figure} 
\centerline{\epsfverbosetrue\epsfxsize=10.5 cm\epsfysize=8.5cm
\epsfbox{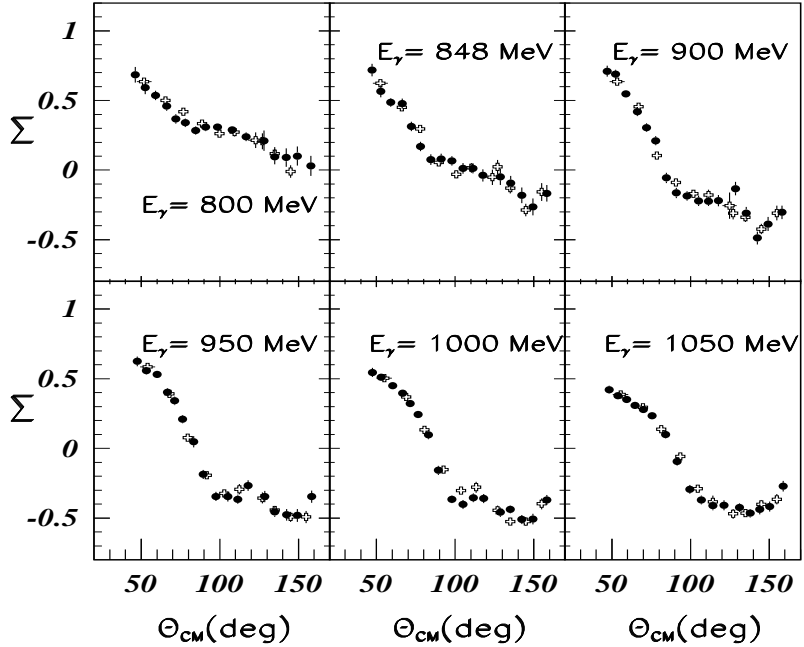}} 
\vspace*{.4cm} 
\caption{Comparison of new and  previous GRAAL results.
 Black circles are the new points, crosses are the results
 of the previous measurement \protect\cite{pin}.}
\label{Figure2}
\end{figure} 

The TOF resolution for neutrons in the shower wall is 
700~ps (FWHM) \cite{rw}.  For fast neutrons in the 
forward direction, this results in a poor energy 
resolution of about 30\% depending on the neutron 
kinetic energy.  The response of the BGO ball for both 
neutrons and pions is not uniquely related to the 
kinetic energy due to the complicated interaction of 
these particles in the detector volume \cite{bgo}.  By 
contrast, the  angles of outgoing particles are measured 
with good resolution: $3^{\circ}$ (FWHM) for pions, 
$2.5-3^{\circ}$ (FWHM) for neutrons in the forward 
direction and $6-8^{\circ}$ (FWHM) for neutrons at 
central angles. For this reason, only the angular 
quantities have been exploited for the identification of 
the $\pi^+n$ final state.  For those events which have 
generated two hits in the detection system, associated 
with a pion and a neutron, the correlation between the 
pion and neutron $\Theta$ angles and the coplanarity 
have been used to select the $\pi^+n$ events.  
After this selection, the spectrum of the reaction 
coplanarity $(\phi_n-\phi_{\pi})$ exhibited 
a narrow peak (Fig.~\ref{Figure1}) lying on the broad
background originating mostly from multipion production.
The contribution of this background, which  never exceeded 3\%,
was  evaluated from the tails in $(\phi_n-\phi_{\pi})$ distributions 
and subtracted separately in each $\Delta E \Delta \Theta \Delta \phi$ bin,
in order to account for the possible azimuthal anisotropy
of this background.

The beam asymmetry $\Sigma$ was extracted from the azimuthal
distribution of selected events for the linearly
polarized beam, normalized to the azimuthal distribution
corresponding to an unpolarized beam,
as it is described in details in \cite{pin}. 
This procedure reduces significantly systematic errors
of the extracted asymmetries. The remaining systematic errors,
estimated as not more than 0.02, originate from the unsignificant
uncertainties in the background subtraction and from the slight variations
of the beam profile on the target. 
Another source of the systematic uncertainty is the possible deterioration
of the laser light polarization on the  mirrors, lenses and the window
of the laser focusing system. This factor adds the relative error
of 1\%.

The present set of 237 $\Sigma$ beam asymmetries, 
measured over the energy and angular ranges of 
800$-$1500~MeV and $40 - 160^{\circ}$, is an extension 
of a previous GRAAL data set consisting of 92 data 
points from 600 to 1050~MeV.  The two sets of points 
are shown at overlapping energies in Fig.~\ref{Figure2}. 
The new data have been measured using the UV laser, 
producing a different beam spectrum and a different 
polarization for each beam energy \cite{gra}.  Some 
improvements in the apparatus and the analysis procedure 
have also been implemented.  Given these differences, 
the reproducibility of our results is excellent and 
supports the quality of both data sets.

\begin{figure} 
\vspace*{-.1cm} 
\centerline{\epsfverbosetrue\epsfxsize=10.5 cm\epsfysize=12.0cm
\epsfbox{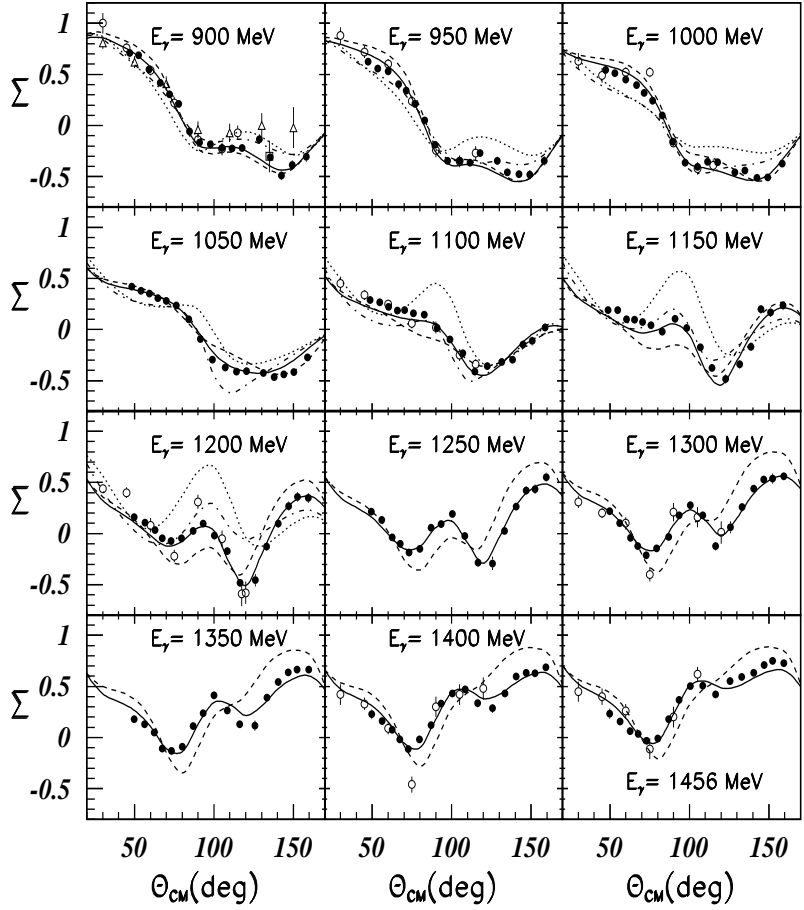}} 
\vspace*{.3cm} 
\caption{$\Sigma$ beam asymmetry at different energies.
 Black circles indicate our results (error bars are the quadratutes
 of the statistic and systematic errors);
 open circles indicate the results of the Daresbury group
 \protect\cite{dar}; open triangles and squares indicate
 the results from SLAC \protect\cite{sl1,sl2}. The solid
 and dashed lines indicate the FA01 and WI00 solutions of
 \protect\cite{str}, respectively; the dotted line is the
 prediction of MAID2000 \protect\cite{maid}; the dashed line
 is MAID2000 after fitting the benchmark database 
 \protect\cite{kam}.}
\label{Figure3}
\end{figure} 

The major part of our new results is shown in 
Fig.~\ref{Figure3}, along with results from the most 
accurate older experiments \cite{dar,sl1,sl2}.  In 
general, our values are in agreement with the previous 
measurements. However, above 950~MeV, we do not confirm 
four points from \cite{dar} at  $75^{\circ}$.  136 data 
points were produced in an almost unexplored domain 
above 1050~MeV, where only 45 old points of lower 
accuracy were available.  The new results also cover 
backward angles above 120$^{\circ}$, where no previous 
measurements exist.  Due to improved angular resolution, 
these results  have been obtained with narrow angular 
bins of $6-10^{\circ}$, which reveal a complicated
angular variation in the $\Sigma$ asymmetry.
 
We have compared our results with the prediction of a 
unitary isobar model of Drechsel \textit{et al.} 
\cite{maid}.  This model, known as MAID2000, in addition 
to Born terms and vector meson exchange, includes all 
3$-$ and 4$-$star resonances to a CM energy of 1800~MeV, 
apart from the $P_{33}(1600)$ and $D_{15}(1675)$.
The latest version of these calculations\cite{kam}, 
which includes resonance parameters obtained from a fit 
to the restricted benchmark data base\cite{bench}, in 
general reproduces our data up to 1200~MeV, though in 
some regions discrepancies are still significant.

In order to assess the impact of our new data, we 
employed the partial wave analysis of the Data Analysis 
Center of The George Washington University (SAID) 
\cite{str}.  Two solutions of the SAID analysis are 
shown in Fig.~\ref{Figure3}: WI00, which was produced 
prior to this measurement, however using our previous
results \cite{pin}, and FA01, a solution
developed after adding our new polarized  data and cross section points 
\cite{Bonn} to the data base.  At the lower energies 
between 900 and 1050~MeV, the WI00 solution is found to 
be in agreement with the new data, although differing 
slightly at angles less than $70^{\circ}$.  At the 
higher energies, differences become more apparent and 
only a qualitative agreement is available.  The FA01 
solution fits our data reasonably well, with $\chi^2$/
data of 555/237, compared to an overall $\chi^2$/data 
$\sim 2$ for the full data base (in both WI00 and FA01).

\begin{figure} 
\vspace*{-.0cm} 
\centerline{\epsfverbosetrue\epsfxsize=10.5 cm\epsfysize=9.0cm
\epsfbox{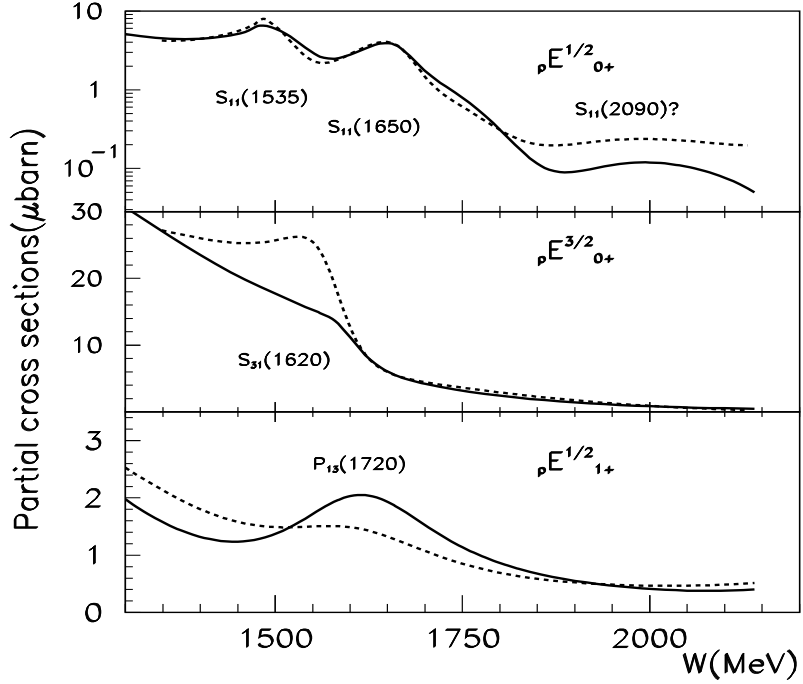}} 
\vspace*{.1cm} 
\caption{$_pE_{0+}^{\frac{1}{2}}$, $_pE_{0+}^{\frac{3}{2}}$, and $_pE_{1+}^{\frac{1}{2}}$ multipoles
 of the SAID analysis.  Solid line gives the FA01 solution,
 and the dashed line is from WI00. }
\label{Figure4}
\end{figure}

In order to illustrate the effect of new data in the SAID 
analysis, we compare 
$_pE_{0+}^{\frac{1}{2}}$, $_pE_{0+}^{\frac{3}{2}}$, and $_pE_{1+}^{\frac{1}{2}}$  
multipoles of the WI00 and FA01 solutions.
Partial cross section contributions from these multipoles 
are shown in Fig.~\ref{Figure4}:\\
{\it $_pE_{0+}^{\frac{1}{2}}$ multipole}. Two distinct peaks 
 correspond to the $S_{11}(1535)$ and $S_{11}(1650)$
 resonances.  The possibility of a third $S_{11}$ 
 resonance, near 1750~MeV, has been considered in 
 several recent studies \cite{said_pin,LiWork,LiSag}.
 The effect of this resonance was found to be 
 particularly significant in a study of eta 
 photoproduction data \cite{LiSag}.  In Fig.~\ref{Figure3}, 
 this state would be hidden in the shoulder of the second
 $S_{11}$ peak of the FA01 solution. 
  Another candidate has been suggested at higher energy 
 (the $S_{11}(2090)$).
 The ``one-star" $S_{11}(2090)$ included in 
 Refs.~\cite{man,plot} has an estimated mass nearer to 
 1900~MeV. A further indication for this state stems from 
 the analysis of low-statistics $\eta '$ photoproduction 
 data \cite{plot}. 

 Differences  appearing in the $_pE_{0+}^{\frac{1}{2}}$
 wave may be indications of 
 such states, which have not been included in either of the 
 displayed fits (WI00 and FA01). However, the signals are weak.  
 New data at higher energies ($\pi^0$ data from GRAAL) are required 
 before any conclusion is possible. These states, if they 
 exist, can only be weakly coupled to $\pi N$ (as is 
 evident in the low cross section). Clearer results may 
 come from other channels, such as $\eta N$, through 
 studies such as the one reported in Ref.~\cite{LiSag}. 
 Forthcoming GRAAL and CLAS $\eta$ photoproduction data
 are clearly of great importance to 
 this issue.\\  
{\it $_pE_{0+}^{\frac{3}{2}}$ multipole}. The peak existing in the
 WI00 solution corresponds to the $S_{31}(1620)$ 
 resonance. This structure nearly vanishes in the FA01 
 solutions. \\
{\it $_pE_{1+}^{\frac{1}{2}}$ multipole}. An opposite trend is found 
 in this multipole. Here a much more pronounced structure, 
 corresponding to the $P_{13}(1720)$, is seen in the 
 revised fit. 
 
At present, the $P_{13}(1720)$ and $S_{31}(1620)$ 
resonances have poorly determined photo-decay amplitudes. 
Neither state can be identified through the ``canonical" 
resonance behavior of its associated multipoles.  A similar 
problem is seen in the analysis of elastic pion-nucleon 
scattering data \cite{said_pin}.  Both states have nearby 
pole-zero pairs in the complex energy plane.  As a result, 
resonance properties are difficult to determine.  
  
The ambiguous situation surrounding the $S_{31}(1620)$ has 
been pointed out in \cite{benh}.  The Particle Data Group 
estimates the photo-decay coupling of the $S_{31}(1620)$  
to be $A_{\frac{1}{2}}=(0.027\pm 0.011)$~GeV$^{-\frac{1}{2}}$ 
\cite{pdg}.  This value was obtained as a weighted average of 
contradicting results from \cite{str,s31}, which range from 
$(-0.026 \pm 0.008)$ to $(0.126 \pm 0.021)$~GeV$^{-\frac{1}{2}}$. 
In a recent multi-channel fit \cite{feu} this quantity was 
found to be near zero.  Our results also suggest a small value.

We have studied the sensitivity of observables in the WI00 and FA01
solutions to the $S_{31}(1620)$ resonance. For W=1620 MeV, we found
an essential difference (10-15\%) between two solutions for 
the $\Sigma$, T, and P onservables in $\pi^+ n$ photoproduction, 
and $\Sigma$ observable in $\pi^0 p$ photoproduction
($\Sigma$, T, P denote beam, target, and recoil asymmetries respectively).
For $\pi^0 p$ T and P observables the difference reaches 30-40\%.
New polarized data from modern photon factories are clearly 
needed to place  more constraints on determination of properties
of this and other resonances.

We are grateful to the staff of the contributing Institutes 
for their assistance in the development and maintenance
of the apparatus. We acknowledge the machine operation 
group of the ESRF for providing stable beam operation. 
Support from the Russian Ministry
of Sciences, INFN Sezione di Roma2, and the Universit\'e Joseph 
Fourier, Grenoble, has been essential for this work.
The GW group gratefully acknowledges a contract from Jefferson 
National Lab under which the work was done.  The Thomas 
Jefferson National Accelerator Facility is operated by the 
Southeastern Universities Research Association under DOE 
contract DE-AC05-84ER40150.  The GW group also acknowledges 
separate support from DOE contract DE-FG02-99ER41110.



\end{document}